\documentclass[useAMS,usenatbib]{mnras}

\usepackage[active]{srcltx}
\usepackage{graphicx}
\usepackage{txfonts}
\usepackage{natbib}
\usepackage{multirow}
\usepackage{longtable}
\usepackage[applemac]{inputenc}
\usepackage{rotating}
\usepackage{tablefootnote}
\usepackage{url}
\usepackage[multiple]{footmisc}
\usepackage{color}
\usepackage{soul}
\definecolor{jaune}{rgb}{1.0, 1.0, 0.0}

\usepackage{caption}
\newcommand{\teff}{$T_{\rm eff}$}

\newcommand{\bz}{\ensuremath{\langle B_z\rangle}}

\newcommand{\kms}{km\,s$^{-1}\,$}
\newcommand{\vsini}{$v\sin i\,$}

\def\gtrsim{\mathrel{\hbox{\rlap{\hbox{\lower4pt\hbox{$\sim$}}}\hbox{$>$}}}}
\def\ltsim{\mathrel{\hbox{\rlap{\hbox{\lower4pt\hbox{$\sim$}}}\hbox{$<$}}}}
\newcommand{\el}{$\varepsilon$~Lupi}

\title[$\varepsilon$ Lupi]{$\varepsilon$ Lupi: measuring the heartbeat of a doubly-magnetic massive binary with BRITE-Constellation}

\author[Herbert Pablo et al.]{H. Pablo\thanks{E-mail: hpablo@aavso.org}$^{1}$, M. Shultz$^{2,3}$, J. Fuller$^{4,5}$, G.A. Wade$^{6}$, E. Paunzen$^{7}$, S. Mathis$^{8,9,10}$,
\newauthor{ J.-B. Le Bouquin$^{11,12}$, A. Pigulski,$^{13}$, G. Handler$^{14}$, E. Alecian$^{11}$, R. Kuschnig$^{15}$,}
\newauthor{A.F.J. Moffat$^{16}$, C. Neiner$^{10}$, A. Popowicz$^{17}$, S. Rucinski$^{18}$, R. Smolec$^{19}$,}
\newauthor{ W. Weiss$^{20}$, K. Zwintz$^{21}$, and the BinaMIcS Collaboration}
\footnotemark[1]\thanks{Based in part on data collected by the BRITE Constellation satellite mission, designed, built, launched, operated and supported by the Austrian Research Promotion Agency (FFG), the University of Vienna, the Technical University of Graz, the University of Innsbruck, the Canadian Space Agency (CSA), the University of Toronto Institute for Aerospace Studies (UTIAS), the Foundation for Polish Science \& Technology (FNiTP MNiSW), and National Science Centre (NCN).}\\
$^{1}$American Association of Variable Star Observers, 49 Bay State Road, Cambridge, MA 02138, USA\\
$^{2}$Annie Jump Cannon Fellow, Department of Physics and Astronomy, University of Delaware, 217 Sharp Lab, Newark, Delaware, 19716, USA\\
$^{3}$Department of Physics and Astronomy, Uppsala University, Box 516, Uppsala 75120 \\
$^{4}$TAPIR, Walter Burke Institute for Theoretical Physics, Mailcode 350-17, Caltech, Pasadena, CA 91125, USA\\
$^{5}$Kavli Institute for Theoretical Physics, Kohn Hall, University of California, Santa Barbara, CA 93106, USA\\
$^{6}$Royal Military College of Canada, Dept. of Physics and Space Science, Kingston, Ontario, K7K 7B4, Canada \\ 
$^{7}$ Department of Theoretical Physics and Astrophysics, Masaryk University, Kotl\'a\v rsk\'a 2, CZ-611 37 Brno, Czech Republic\\
$^{8}$ IRFU, CEA, Universit\'e Paris-Saclay, F-91191 Gif-sur-Yvette, France\\
$^{9}$ Universit\'e Paris Diderot, AIM, Sorbonne Paris Cit\'e, CEA, CNRS, F-91191 Gif-sur-Yvette, France\\
$^{10}$ LESIA, Observatoire de Paris, PSL University, CNRS, Sorbonne Universit\'{e}, Universit\'{e} de Paris, 5 place\\ Jules Janssen, F-92195 Meudon, France\\
$^{11}$ Universit\'e Grenoble Alpes, CNRS, IPAG, 38000, Grenoble, France\\
$^{12}$ Department of Astronomy, University of Michigan, Ann Arbor, MI 48109, USA\\
$^{13}$ Instytut Astronomiczny, Uniwersytet Wroc{\l}awski, Kopernika 11, 51-622 Wroc{\l}aw, Poland\\
$^{14}$ Centrum Astronomiczne im.~M.\,Kopernika, Polska Akademia Nauk, Bartycka 18, 00-716 Warszawa, Poland\\
$^{15}$ Graz University of Technology, Institute of Communication Networks and Satellite Communications, Inffeldgasse 12, 8010 Graz, Austria\\
$^{16}$ D\'epartement de physique and Centre de Recherche en Astrophysique du Qu\'ebec (CRAQ), Universit\'e de Montr\'eal, C.P. 6128,\\ Succ.~Centre-Ville, Montr\'eal, Qu\'ebec, H3C 3J7, Canada\\
$^{17}$ Silesian University of Technology, Institute of Automatic Control, Gliwice, Akademicka 16, Poland\\
$^{18}$Dept. of Astronomy and Astrophysics, University of Toronto, 50 St George Street, Toronto, ON M5S 3H4, Canada\\
$^{19}$ Nicolaus Copernicus Astronomical Center, Bartycka 18, 00-716 Warszawa, Poland \\
$^{20}$ Institut f{\"u}r Astrophysik, Universit{\"a}t Wien, T{\"u}rkenschanzstrasse 17, 1180 Wien, Austria\\
$^{21}$ Universit{\"a}t Innsbruck, Institut f{\"u}r Astro- und Teilchenphysik Technikerstrasse 25/8, A-6020 Innsbruck\\
}

\begin{document}

\date{Accepted . Received , in original form }

\pagerange{\pageref{firstpage}--\pageref{lastpage}} \pubyear{2019}

\maketitle

\label{firstpage}

\begin{abstract}
\el\ A is a binary system consisting of two main sequence early B-type stars Aa and Ab in a  short period, moderately eccentric orbit. The close binary pair is the only doubly-magnetic massive binary currently known. Using photometric data from the BRITE-Constellation we identify a modest heartbeat variation. Combining the photometry with radial velocities of both components we determine a full orbital solution including empirical masses and radii. These results are compared with stellar evolution models as well as interferometry and the differences discussed. We also find additional photometric variability at several frequencies, finding it unlikely these frequencies can be caused by tidally excited oscillations. We do, however, determine that these signals are consistent with gravity mode pulsations typical for slowly pulsating B stars.  Finally we discuss how the evolution of this system will be affected by magnetism, determining that tidal interactions will still be dominant.     
\end{abstract}


\section{Introduction}
\el\ is a multiple system consisting of \el\ Aa and Ab, a double lined spectroscopic binary (SB) with two early B-type components and a distant tertiary companion \el\ B with a separation of $\approx0.2''$. The orbit of the A-B system, first derived by \cite{Zirm07}, has a period of ~740 yrs. \el\ B\footnote{The A-B system is often referred to in the literature as COP 2.} was initially resolved in 1883 by Ralph Copeland \citep{elb97}. However, outside of relative light contribution, little is known about the B component specifically.  \el\ A (heretofore referred to as \el), by contrast, has been well studied, with first evidence of its binarity found by \cite{moore11}. It was later confirmed to be a double-lined SB by \cite{thack70} with a period of 4.56 d and component spectral types identified as B3IV and B3V. 

Modern analyses of the components' spectra  and their RV motions yield results in qualitative agreement with those of \citet{thack70}: \el\ is a B2V/B3V SB2 in an eccentric ($e=0.28$) orbit with a period of 4.56~d \citep[e.g.][]{2005A&A...440..249U}.  Some evidence has been presented that one or both of the components exhibits $\beta$~Cep pulsations \citep{2005A&A...440..249U}.

\citet{2009AN....330..317H} reported a magnetic field in the combined spectrum of this system, a result subsequently confirmed by \citet{2012ApJ...750....2S}. Using deep, high-resolution ESPADOnS spectropolarimetry \el\ acquired as part of the Binarity and Magnetic Interactions in various classes of Stars (BinaMIcS) survey \citep{binamics}, \citet{2015MNRAS.454L...1S} reported the presence of surface magnetic fields in the primary and secondary of 200 G and 100 G, respectively.  As the only known ``doubly-magnetic'' early-type binary system \el\ is a unique object of particular interest for understanding the origin and evolution of the magnetic fields of hot stars and their interactions in such compact systems \citep[e.g.][]{MathisBinamics}.

What has been absent in the study of the \el\ system is time-series photometry. The nature of this system makes it likely to exhibit variable broadband flux, potentially on a variety of timescales. The components are suspected to be $\beta$~Cep pulsators, corresponding to variability on a timescale of hours \citep[Please see][for more information on $\beta$~Cep pulsators]{1955PASP...67..135S,1978ARA&A..16..215L,1993SSRv...62...95S}. In some B-type stars magnetic fields stabilise atmospheric motions and allow the accumulation of non-uniform abundance distributions of various chemical elements \citep[e.g.][]{1981A&A...101...16A,2011MNRAS.418..986A,2015MNRAS.454.3143A,2016MNRAS.457...74S}. Such structures commonly introduce photometric variability modulated by rotation \citep[e.g.][]{bp1, bp2,2019arXiv190200326J}; if chemical spots are present on \el\ this modulation would likely be on a timescale of days \citep{2005A&A...440..249U}. Magnetic fields may also channel radiatively-driven stellar winds, confining wind plasma to produce co-rotating magnetospheres \citep[e.g.][]{2002ApJ...576..413U,town05}. The magnetospheric plasma may occult the star and scatter starlight, modulating the systemic brightness on the rotational timescale \citep[e.g.][]{town2008,town2013}. Magnetospheres can also be dramatically altered by orbital interactions \citep{2018MNRAS.475..839S}. Finally, in a close binary system like \el, photometric variability on  orbital timescales is possible, through eclipses, tidal interaction and (potentially) mass and energy transfer effects  \citep[][and references therein]{fuller:12} \footnote{In fact, the (variable) proximity of the two components of \el\ led \citet{2015MNRAS.454L...1S} to speculate that their magnetospheres and overlapping Alfven radii may undergo reconnection events during their orbit.}.

In this paper, we provide an in depth analysis of the photometric variability present in the \el\ system, using data mainly from the BRIght Target Explorer (BRITE) Constellation \citep{briteI}. We start by characterizing \el\ B and understanding its contribution to the overall flux of the system (Sect. 3). We follow this by characterizing the source of the orbital variation seen, determining it is due to the heartbeat effect. Using this heartbeat in combination with radial velocities we produce a full empirical solution of the \el\ A system and compare these results with interferometric and stellar evolution models (Sect. 4). Then we undertake a full frequency analysis, discussing the role tidal oscillations may play in this system (Sect. 5). Finally, we explore the effect that magnetism could have had in the system's evolution (Sect. 6).

\section{Observations}

\subsection{Photometry}

Nine months of high-precision, two-colour photometry of $\varepsilon$ Lupi were obtained by the BRITE constellation of satellites during two separate observational campaigns between March of 2014 and August of 2015 by UniBRITE (UBr), BRITE-Austria (BAb), BRITE Lem (BLb), and BRITE Toronto (BTr) (see Tab.~\ref{tab:observations} for a complete summary).  The raw images from the satellites were processed using the pipeline outlined by \cite{pop16} and \cite{briteiii}.  While the pipeline effectively mitigates many of the issues associated with BRITE photometry such as hot pixels \citep{BriteII},  the reduced photometry remains strongly imprinted with instrumental trends.  These systematics are effectively identified and removed via decorrelation, i.e. correcting the dependence of the measured flux on different instrumental parameters such as CCD temperature and $x/y$ position of the stellar profile within the raster. This was carried out in a manner similar to that outlined by \citet{pigu16}. An additional decorrelation mitigating the impact of temperature on the point-spread function (PSF) was applied according to the procedure outlined by \cite{bram17}. Finally the data were then binned on the satellite's orbital period and the RMS errors calculated within each orbit. As all the frequencies we are concerned with are well below the Nyquist frequency of the binned data ($\approx 7 \rm{d}^{-1}$) this allows us to reduce scatter and achieve smaller uncertainties.  The average RMS error per satellite orbital mean, after processing, are reported for each satellite in Tab.~\ref{tab:observations}.

\begin{table}

\caption{BRITE observations of $\varepsilon$ Lupi. The first two capital letters of the satellite moniker represent the satellite name, UniBRITE (UB), BRITE-Austria (BA),  BRITE-Lem (BL), and BRITE-Toronto (BT), while the last lower case letter represents the filter, red (r) or blue (b).  The quoted error (RMS) is per satellite orbital mean, in parts per thousand (ppt).}\label{tab:observations}
	\resizebox{\columnwidth}{!}{\begin{tabular}{lcccc }
	\hline    
    Field Name & Satellite & Duration (d) & Duty Cycle (\%) & RMS (ppt) \\
	\hline
     Centaurus I & UBr &  145 & 67  & 1.84 \\
            & BAb & 131 & 25 & 1.62 \\
            & BTr & 6 &  64 & 0.41 \\
	Scorpius I & BLb & 117 & 60 & 5.1 \\
	\hline	
    \hline	            
    \end{tabular}}
\end{table}

Additional corrections were applied on a per dataset basis to further enhance analysis efforts. All datasets had long term trends removed using a LOWESS filter \citep{cleveland79} as well as sigma clipping (typically $3 \sigma$) to remove strong outliers. Data from both blue filter satellites required more attention, which is detailed in Sect.~\ref{sect:model}. Finally, while the BTr data are of the highest quality, they were taken only as part of a commissioning run and the short baseline is sufficient for neither binary nor frequency analysis. 

In addition, we also made use of data from the Solar Mass Ejection Imager (SMEI) \citep{smei2004} spanning the years 2003 to 2010.  While the errors on the individual SMEI data points are much higher than BRITE, with an average point to point error of $\approx 9 $ parts per thousand (ppt), the much longer time baseline makes the data valuable for frequency determination and stability considerations.  Therefore we use the SMEI data to supplement BRITE in our Fourier analysis.

\subsection{ESPaDOnS spectroscopy}

ESPaDOnS is a high-resolution ($\lambda/\Delta\lambda\sim65,000$) spectropolarimeter installed at the Canada-France-Hawaii Telescope (CFHT). It acquires echelle spectra covering the spectral region from 369.3 to 1048 nm across 40 spectral orders. A spectropolarimetric observation consists of 4 sub-exposures corresponding to different orientations of the instrument's polarimetric optics, yielding 4 unpolarized intensity (Stokes $I$) spectra, one circularly polarized (Stokes $V$) spectrum, and two diagnostic null ($N$) spectra. The reduction of ESPaDOnS data and extraction of the polarized spectra via the CFHT's dedicated Libre-Esprit+Upena pipeline is discussed in detail by \cite{2016MNRAS.456....2W}.

Observations of $\varepsilon$ Lupi were acquired by the Magnetism in Massive Stars (MiMeS); \citep{2016MNRAS.456....2W} and Binarity and Magnetic interactions in various classes of Stars (BinaMIcS); \citep{binamics} Large Programs, as well as a PI program\footnote{Program code CFHT14AC010.}. A total of 91 Stokes $V$ spectropolarimetric sequences were obtained on 14 nights, with between 4 and 11 sequences obtained each night. Of these, the data acquired during 10 nights have been published \citep{2015MNRAS.454L...1S}. The 4 additional nights of data were obtained in the same fashion as the observation obtained on 2015-04-09, i.e.\ several closely spaced observations were obtained using a uniform sub-exposure time of 65 s, spanning approximately 1 hr. Due to the close temporal spacing of these data, there is essentially no radial velocity (RV) variation between sub-exposures. Therefore the data obtained during each night were combined to yield a single high-quality spectrum with a peak per-pixel signal-to-noise ratio S/N$\sim$5000 in the final co-added Stokes $V$ spectra.  

RVs were measured using a two-step process identical to that of \cite{2017MNRAS.465.2517W}. First, synthetic two-component line profiles were fit to the Si~{\sc iii} 455.3 nm line using the tool described by \cite{2017MNRAS.465.2432G}. In the second step, the first set of RVs were used to initialize the disentangling of the Si~{\sc iii} 455.3 nm line profiles using essentially the method outlined in \cite{2006AA...448..283G}, with the refinement that the RVs were re-measured at each iteration of the disentangling process using the centre-of-gravity method. Uncertainties were estimated to be $\sim 5$~\kms. The measured ESPaDOnS RVs are given in Table \ref{rv_tab}. 

\begin{table}
\centering
\caption[RV measurments]{Radial velocity (RV) measurements from ESPaDOnS data. RV$_{\rm P}$ and RV$_{\rm S}$ are velocities of the primary and secondary component respectively.  $\phi_{\rm orb}$ are computed using the ephemeris corresponding to the 3 LC solution (see Table \protect\ref{tab:binary_params}). Uncertainties are estimated to be about 5 \kms.}

\begin{tabular}{lllrr}
\hline
HJD      & Date & $\phi_{\rm orb}$ & RV$_{\rm P}$ & RV$_{\rm S}$ \\
$-2450000$ &  &                  & (\kms)       & (\kms)       \\
\hline
5634.15385 & 2011-03-13 & 0.8116 & $-13.0$ &  $11.0$ \\
5727.81171 & 2011-06-15 & 0.3522 & $-24.9$ &  $30.9$ \\
6756.96216 & 2014-04-09 & 0.0606 &  $61.9$ & $-76.6$ \\
6760.95725 & 2014-04-13 & 0.9368 &  $25.0$ & $-30.8$ \\
6816.85240 & 2014-06-08 & 0.1954 &  $37.2$ & $-46.1$ \\
6821.87067 & 2014-06-13 & 0.2960 &  $-9.8$ &  $12.1$ \\
6822.89135 & 2014-06-14 & 0.5199 & $-42.6$ &  $48.3$ \\
6824.87508 & 2014-06-16 & 0.9549 &  $30.7$ & $-36.2$ \\
6819.86772 & 2014-06-11 & 0.8567 &   $3.0$ &  $-2.2$ \\
7122.02504 & 2015-04-09 & 0.1245 &  $64.2$ & $-79.4$ \\
7199.81289 & 2015-06-26 & 0.1845 &  $45.0$ & $-50.0$ \\
7201.81305 & 2015-06-28 & 0.6232 & $-35.0$ &  $45.0$ \\
7228.75673 & 2015-07-25 & 0.5323 & $-41.3$ &  $48.0$ \\
7230.78974 & 2015-07-27 & 0.9782 &  $38.5$ & $-43.9$ \\
\hline
\hline
\end{tabular}
\label{rv_tab}
\end{table}

\subsection{VLTI interferometry}\label{subsect:vlti}

\el\ was observed by the PIONIER\footnote{http://ipag.osug.fr/pionier} instrument from the Very Large Telescope Interferometer \citep{Haguenauer:2010} on 2014-06-09 and 2014-06-10.PIONIER combined the four 1.8m Auxiliary Telescopes in the $H$-band with a spectral resolution of R=15. Each observation consists of the standard Calibration-Science-Calibration sequence. On such a bright target, the SNR ($\sim$50) is limited by the calibration accuracy even with the short 5min integration on target. Data were reduced and calibrated with the \texttt{pndrs}\footnote{http://www.jmmc.fr/pndrs} package \citep[][]{Le-Bouquin:2011}. These observations spatially resolved  the spectroscopic binary and are listed in Table \ref{vlti_tab}.

\section{\el\ B}\label{epslupb}

In order to characterize the inner binary it is important to understand contributions from the tertiary component \el\ B. The B companion contributes to 20\% of the $V$-band flux \citep{2013MNRAS.436.1694R}) and is expected to be at $\approx$200~mas at the time of the PIONIER observation\citep{2015yCat....102026M}. At this separation, it is expected to appear as a fully resolved component because it is wider than the interferometric field of view (FOV) (about 50~mas, defined by the baseline lengths and our low spectral resolution) but still within the photometric FOV as defined by \citet{2017A&A...597A...8D} (about 250~mas, defined by the diffraction of the 1.8m telescopes). Our PIONIER interferometric observations indeed clearly reveal a third, fully resolved component contributing $17 \pm 3$\% of the total $H$-band flux (see Sect.5.3). This is curious as such a significant contribution should be detectable in the stellar spectrum, yet no such signature has been reported (e.g. \citealt{2005A&A...440..249U}).

\begin{figure}
\begin{center}
\includegraphics[width=7cm]{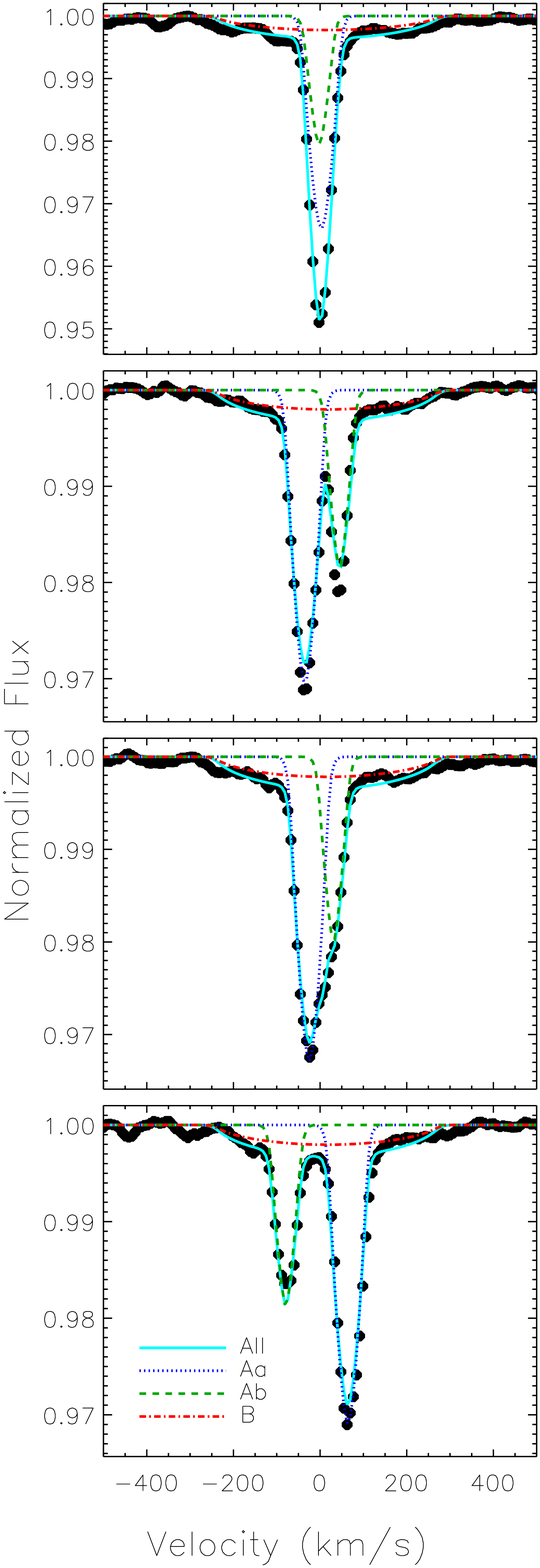}
\caption{LSD intensity profiles extracted from ESPaDOnS data (black circles) and fit with a 3-component model (lines).}
\label{3starfit}
\end{center}
\end{figure}

The high flux ratio of the B component motivated a re-examination of the ESPaDOnS dataset. One possible explanation for the failure to detect $\varepsilon$ Lupi B would be if the star has a very large \vsini, since in this case its line profile would be difficult to distinguish from the continuum. To explore this possibility, Least-Squares Deconvolution (LSD) profiles were extracted using a custom atomic line mask from the Vienna Atomic Line Database \citep[VALD3;][]{1995A&AS..112..525P, ryabchikova1997,1999A&AS..138..119K,2000BaltA...9..590K,Ryabchikova2015} using a line list extracted for a 20 kK, $\log{g}=4.0$ star. The line mask was prepared by removing all lines overlapping with the wings of H or He lines; lines in spectral regions contaminated with telluric lines, and lines with depths less than 10\% below the continuum. 

LSD profiles were extracted using a velocity range of $\pm 600$~\kms, and a velocity pixel of $7.2$~\kms. Fig.~\ref{3starfit} shows 4 representative LSD profiles with orbital phases matching as closely as possible 0.0, 0.25, 0.5, and 0.75. There is a clear depression in the continuum between approximately $\pm~250$~\kms. This depression was not noticed before as the velocity width of $\pm$300~\kms used by \cite{2012ApJ...750....2S,2015MNRAS.454L...1S} is similar to the width of the depression. 

The LSD profiles were fit with a 3-component model using the parameterized routine described by \cite{2017MNRAS.465.2432G}, which simultaneously determines \vsini, RV, and EW ratios of the line profiles. Fig.\ \ref{3starfit} shows fits for Aa, Ab, B, and the combined line profiles using the best-fit parameters from the full ESPaDOnS dataset. The routine yielded \vsini~$=260 \pm 12$~\kms~for the broad-lined component, and indicated that it contributes about 25\% of the total EW of the line profile. The EW ratio is compatible with the expected flux ratio from interferometry and photometry, suggesting that the broad-lined component is indeed the spectroscopic signature of $\varepsilon$ Lupi B. The broad-lined component's RV is $ 16 \pm 9$~\kms, and is consistent with no variation over both short and long timescales. 

An attempt was made to isolate the contribution of the broad-lined star using line masks optimized for lower \teff. However in all cases the results were inferior to those obtained with the 20 kK line mask. This suggests that the B component in fact has a \teff~similar to that of the Ab component, as would be expected if it contributes a similar amount of flux to the total light. This would make the star a early B star. The very high \vsini~could be compatible with a classical Be star. However, as there is no emission detectable in H$\alpha$ or any other line, nor has emission ever been reported, it is more likely a Bn star, i.e.\ a non-emission line, rapidly-rotating B-type star similar to a classical Be star \citep[see e.g.][]{2013A&ARv..21...69R}.

The very high \vsini~of $\varepsilon$ Lupi B suggests that its rotational axis is not aligned with the orbital axis of the Aab system, since its radius is presumably similar to that of $\varepsilon$ Lupi Ab. Given that the star is a wide binary, with an orbit of several decades, this is not unexpected. 

While $\varepsilon$ Lupi B contributes a significant amount of light to the integrated flux, its contribution to the EW within the line profiles of Aa and Ab is very small. Even so, the \teff~measurements made by \cite{2019arXiv190202713S} were revisited. EWs for the individual components were made by fitting the Si~{\sc ii} 634.7 nm and Si~{\sc iii} 455.3 nm line profiles of the components with a 3-star model using the paramaterized line profile fitting package described by \cite{2017MNRAS.465.2432G}. The \teff~of each component was then determined from BSTAR2006 synthetic spectra, as described by \cite{2019arXiv190202713S}, yielding \teff$_{\rm Aa} = 21 \pm 1$ kK, \teff$_{\rm Ab} = 19 \pm 1$ kK, and \teff$_{\rm B} = 18 \pm 2$ kK, i.e.\ the results for the Aab components are essentially unchanged from those obtained assuming a 2-star model. The result for the B component should however be interpreted with caution, as while its contribution to Si~{\sc ii} is quite obvious, it is almost undetectable in Si~{\sc iii}.

The contribution of $\varepsilon$ Lupi B to the integrated light in the BRITE blue and red filters was estimated using synthetic spectra from the NLTE BSTAR2006 library \citep{lanzhubeny2007}. Since these spectra extend only to 1000 nm, but the interferometric flux ratios are in the $H$-Band with an effective wavelength of 1630 nm, we first used Planck functions to estimate the corrections for the different stellar radii in the $H$-Band by normalizing the respective Planck functions to the observed flux ratios. These radii corrections were then applied to BSTAR2006 SEDs to obtain the flux ratios in the BRITE blue and red filters, using \teff~$=21$ kK for Aa, 19 kK for Ab, and 15 to 19 kK for B. The result is that the B component contributes approximately 15\% of the light in the blue and red bands, essentially independent of wavelength within this range. Propagation of the uncertainties in the interferometric flux ratios yields an uncertainty of $\sim 10\%$ of the total light, which is larger than the likely systematic uncertainty from utilizing Planck functions in the $H$-Band. 

The longitudinal magnetic field measurements \bz~performed by \cite{2015MNRAS.454L...1S,2018MNRAS.475.5144S}, and used to estimate the Aab components' surface magnetic dipole strengths, should not be affected by the third light. The very large \vsini~of the B component means that the Aab are blended at all phases in a similar fashion, with the effect equivalent to the continuum being depressed. In principle this slightly increases the EW of Stokes $I$, possibly leading to an underestimation of \bz~and, hence $B_{\rm d}$. However, as described by \cite{2018MNRAS.475.5144S}, the Stokes $I$ LSD profiles were renormalized to the local continuum outside the line profiles before measuring \bz; thus, the B component's influence was automatically accounted for.

\section{Characterization of Photometric Variability}

\begin{figure}
\includegraphics[scale=0.3]{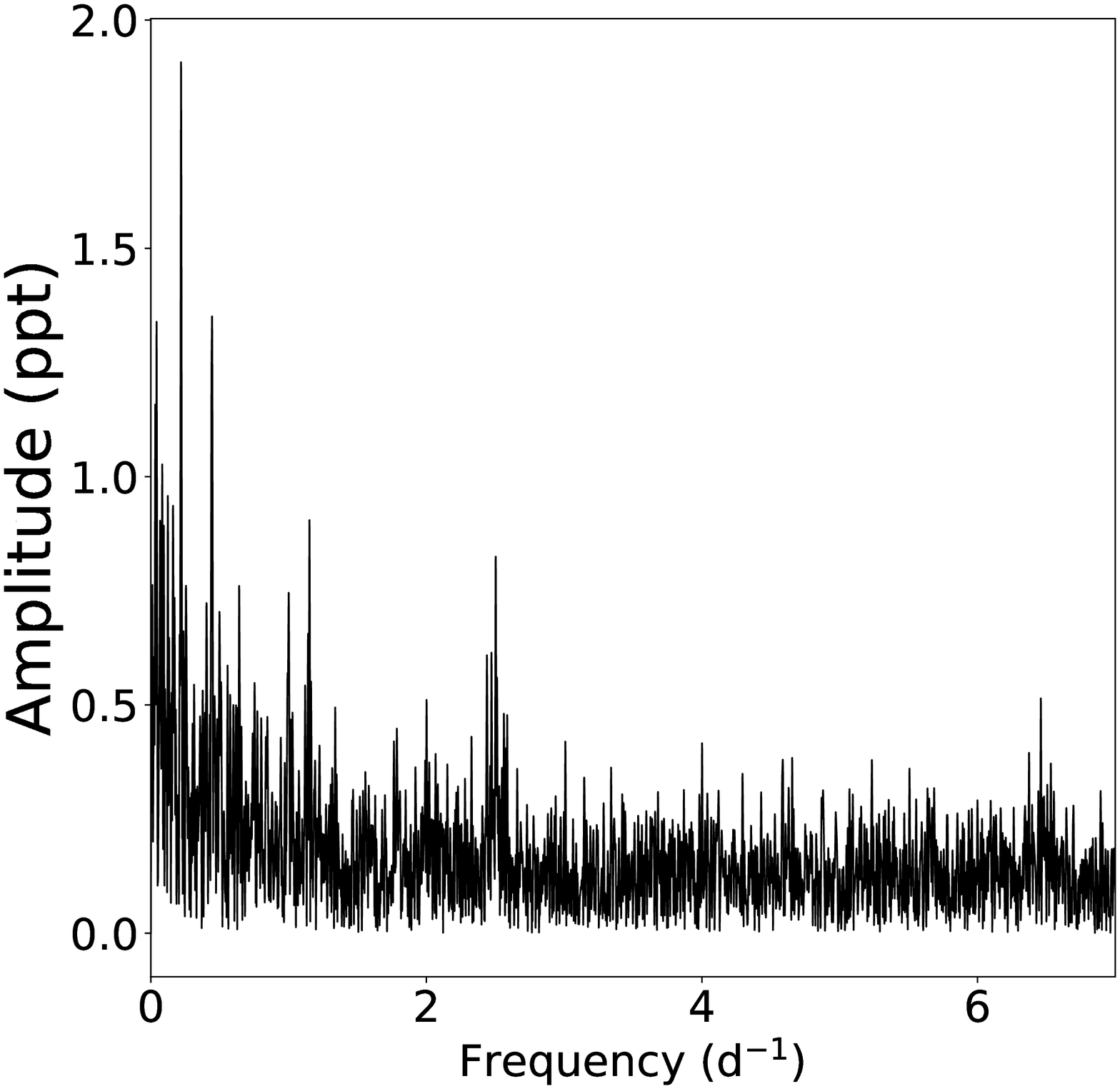}
\caption{Point-error-weighted FS of the full, unfiltered, UBr dataset of \el.  The largest peak occurs at $0.219~\rm{d}^{-1}$, the orbital frequency of the binary.}
\label{full_ft}
\end{figure}

Initial inspection of the BRITE light curves showed significant variability, though the nature of this variation was not easily identifiable. A frequency spectrum (FS) of these data (see Figure~\ref{full_ft}), shows several well defined peaks, the highest occurring at the known orbital frequency. The existence of this peak, while not surprising, had never before been seen, likely due to its small amplitude. Phasing of the photometry on the binary period shows  non-sinusoidal periodic variability (see top panel of Figure~\ref{fig:bin_fit}). 

A critical piece of analysis for this system is determining the origin of this variability. While there are various potential origins capable of explaining this variability, we will focus on the three most probable: magnetism (and rotational modulation), Doppler boosting, and binarity. 

Since the expected magnetic effects are tied to rotation (e.g. surface chemical abundance spots), the system must be synchronized for magnetism to be the dominant variability source since it is commensurate with the system's orbital period. While our spectroscopic data do not have the precision necessary to empirically determine the rotation rates of the Aa and Ab components, we are able to estimate the synchronization timescale ($t_{sync}$). For stars with radiative envelopes like the components of \el,  $t_{sync}$ can be estimated following formula 4.28 from \cite{zahn75}. Using the known binary parameters  of \el~and typical stellar models for both a 7 $M_{\odot}$ and 10 $M_{\odot}$ primary \citep[also from][]{zahn75}  we were able to obtain reasonable limits on $t_{sync}$, placing it between 3.6 $\times 10^{7}$ and 7.3 $\times 10^{7}$ yrs.  While this range is relatively large, the lower limit is still above the age estimate we derived for the system  $ \approx 1.8 \times 10^{7}$ yrs (see Section~\ref{sect:evol_model} for more information) making synchronization unlikely. 

Next we explored the idea of Doppler boosting, a relativistic change in the output of light as a function of the stars' movement in our line of sight. The order of this effect can be calculated, following the procedure outlined by \cite{loeb03}. While this effect for each star would trace the shape of its velocity curve, the individual components are photometrically indistinguishable and the true shape for the system would instead follow the form created by the addition of the two curves. The variability caused by boosting does mimic the light curve variability seen in \el~. However, since the two components are similar in mass the calculated amplitude is roughly $1\times10^{-5}$ in flux, much too small to account for observed amplitude which is $\approx 5\times10^{-3}$. 

Finally we explored binarity. The shape of the phase-folded curve which shows increased amplitude at periastron is in line with the heartbeat phenomenon. This heartbeat is a result of tidal distortion and is normally observed in systems with high eccentricity as the effect is highly dependent on the distance between the two components \citep{heartbeats}. Since this variability is strongly dependent on orbital parameters, most notably inclination, its presence allows for  determination of masses and radii without the need for eclipses. While this effect has been seen in two other massive binaries \citep{pablo:17,2019arXiv190100005J}, the modest eccentricity of the \el~system makes for a less than ideal candidate to show this effect. However, as will be demonstrated in  Sect.~\ref{sect:model}, the heartbeat phenomenon not only describes the light curve variability well, but it is also consistent with the parameters determined from radial-velocity measurements.

\section{A Binary Solution to \el\ A}

The determination of fundamental parameters is a key to understanding such a unique binary system. Below we use photometric, spectroscopic, and interferometric measurements to find empirical values of the system's parameters and discuss discrepancies between the results of the different methods used.

\subsection{Binary modelling}
\label{sect:model}

Due to the long history of observations of \el\ we have a wealth of radial velocity data available. However, using datasets from different epochs and telescopes requires some initial preparation. Our RV data come from three distinct sources and time periods: \cite{thack70}, \cite{2005A&A...440..249U} data from 2003, and ESPaDOnS measurements from 2011 to 2015. First, all time points without a measurement from both components were removed. These values are often suspect as it implies lines from the primary and secondary were blended. Next, the data were split by epoch to avoid issues of apsidal motion. Finally, the 2003 data were split even further by observatory. Our goal now is two-fold. First we need to remove any difference in the systemic velocity, as our binary simulation program cannot account for changes over time. Second we need well determined errors that are consistent across all datasets. The first step is to achieve consistency within a given epoch. To this end, an orbital fit was applied to the \cite{2005A&A...440..249U} data from both observatories individually using least squares and a discrepancy of 2.5 km/s in the systemic velocity was found and removed. Then the same fitting procedure was applied to each epoch and the errors were re-evaluated for each dataset based on the scatter of the residuals. Since this fit is dependent on the error values chosen, the fit was repeated, typically 3 to 4 times, until there was noticeable change in the value of the errors. Finally, the three epochs were adjusted to have the same systemic velocity, a maximum change of roughly 8 km/s. 

The BRITE light curves also required specialized treatment due to the low amplitude of the signal relative to the RMS error. First, for all three datasets, trends longer than the binary period were removed using a LOWESS filter \citep{lowess}. While this was sufficient in removing all long term variation for UBr, both blue satellites required more attention. BAb, despite data being taken in the same run as UBr, struggled to lock on to the target field resulting in two small observation windows of 36 and 18 days with a gap of over 75 days in between. Even after LOWESS filtering, the first window had several small gaps and accompanying discontinuities which dominate the variability. Therefore only the last 18 days were used in the fit. The BLb data were centred on a different observing field, resulting in \el\ being on the edge of the CCD and having only a third the number of counts per observation as BAb. Since the resultant low signal-to-noise makes detection of the modest binary signal almost impossible, data quality was of utmost importance. Therefore, each of the 6 setup files (see \citealt{BriteII} for an explanation of the significance of these files), was analyzed individually. One had only about one-fifth of the counts of all other observations and was removed. Each of the remaining setups were checked to see if the scatter was small enough for the binary period to be clearly seen in the FS. This left one nearly continuous 55-day chunk.

\begin{table*}
\begin{center}
\caption{Best fit values for $\varepsilon$ Lupi A system parameters with $\pm$ 2 $\sigma$ error bars, along side those of Uytterhoven et al. (2005).  $T_{0}$ from that work has been adjusted slightly at to be in the  same orbital cycle as our result for ease of comparison.  $\omega$ quoted is relative to $T_{0}$.} \label{tab:binary_params} 
	\begin{tabular}{r c c c c c l c}
    \hline
	&	& \multicolumn{2}{c}{3 LC Solution}  & \multicolumn{2}{c}{1 LC Solution} & \multicolumn{2}{c}{Uytterhoven et al. (2005)}  \\
	& Parameter & Primary   &   Secondary & Primary & Secondary \\
	\hline
	\vspace{2mm}
	& $P_{\rm orb}$(d) & \multicolumn{2}{c}{ $4.559646_{-8\times10^{-6}}^{5\times10^{-6}} $} & \multicolumn{2}{c}{ $4.559643_{-4\times10^{-6}}^{7\times10^{-6}}$} & \multicolumn{2}{c}{$4.55983 \pm 1\times10^{-5}$}\\
	& $T_{0} (\rm{HJD}-2400000)$ & \multicolumn{2}{c}{$39379.875_{-0.019}^{0.024}$}  & \multicolumn{2}{c}{$39379.883_{-0.025}^{0.016}$}& \multicolumn{2}{c}{$39379.90 \pm 0.05$} \\
	& $i~(^\circ)$  & \multicolumn{2}{c}{$18.8_{-1.4}^{1.6}$} & \multicolumn{2}{c}{$20.2_{-1.9}^{0.7}$} & \multicolumn{2}{c}{--} \\
	\vspace{2mm}
	& $\omega~(^\circ)$ & \multicolumn{2}{c}{$335.7_{-4.5}^{4.8}$} & \multicolumn{2}{c}{$334.0_{-3.5}^{6.5}$} & \multicolumn{2}{c}{$347 \pm 5$} \\
	\vspace{2mm}
    &$\frac{d\omega}{dt}~(^\circ/yr)$ & \multicolumn{2}{c}{$1.1_{-0.1}^{0.1}$} & \multicolumn{2}{c}{$1.2_{-0.2}^{0.1}$} & \multicolumn{2}{c}{$0.8 \pm 0.2$} \\
	& $e$ & \multicolumn{2}{c}{$0.2806_{-0.0047}^{0.0059}$} & \multicolumn{2}{c}{$0.2821_{-0.0039}^{0.0053}$}  & \multicolumn{2}{c}{$0.272 \pm 0.006$} \\
	\vspace{2mm}
	& $q$ & \multicolumn{2}{c}{$0.842_{-0.010}^{0.004}$} & \multicolumn{2}{c}{$0.8393_{-0.0072}^{0.0064}$}  & \multicolumn{2}{c}{$0.84 \pm 0.01 $} \\
	\vspace{2mm}
	& $a~(R_{\odot})$ &  \multicolumn{2}{c}{$31.5_{-2.3}^{2.5} $} & \multicolumn{2}{c}{$29.5_{-1.0}^{2.8}$} & \multicolumn{2}{c}{--}\\
	\vspace{2mm}
	& $v_{\gamma}~(km s^{-1})$ & \multicolumn{2}{c}{$-0.08_{-0.17}^{0.22} $} & \multicolumn{2}{c}{$-0.03_{-0.21}^{0.14}$}  & \multicolumn{2}{c}{$3 \pm 3$} \\
	\vspace{2mm}    
	& $T_{\rm{eff}}~(K)$  & $20500$ (fixed) & 18000 (fixed) &  $20500$ (fixed) & 18000 (fixed) &\multicolumn{2}{c}{-- \hspace{8mm} --} \\
	\vspace{2mm}
	& $R~(R_{\odot})$ & $4.64_{-0.48}^{0.37}$ & $4.83_{-0.46}^{0.42}$ & $4.51_{-0.59}^{0.31}$ & $4.47_{-0.36}^{0.63}$ & \multicolumn{2}{c}{-- \hspace{8mm} --} \\	
	\vspace{2mm}
	\vspace{2mm}
	& $M~(M_{\odot})$ & $11.0_{-2.2}^{2.9}$ & $9.2_{-1.9}^{2.4}$ & $9.0_{-0.9}^{2.9}$ & $7.6_{-0.7}^{2.4}$ & \multicolumn{2}{c}{-- \hspace{8mm} --}\\
	& $L~(L_{\odot})$ & $3407_{-567}^{658}$ & $2197_{-399}^{489}$ & $3225_{-782}^{464}$ & $1883_{-289}^{573}$ & \multicolumn{2}{c}{-- \hspace{8mm} --}\\
	\hline
    \hline
\end{tabular}
\end{center}
\end{table*}

With these reductions made to the data we are able to do a full binary analysis. Because of the disparity in data quality observed between the photometric datasets we found one solution using all 6 datasets (3LCS) mentioned above, and a second with only the UBr in combination with the RV datasets (1LCS) (see Table~\ref{tab:binary_params}). In each case we used the binary simulation program PHOEBE \citep{phoebe} to create models of all included datasets simultaneously. These models were then compared to their respective datasets using  $\chi^{2}$.  The sum of these $\chi^{2}$ was used to create single value which was then minimized using a Monte Carlo Markov chain (MCMC) code to probe the parameter space and determine uncertainties. Our MCMC implementation uses the Python package {\tt emcee} \citep{mcmc}. As there are dozens of possible parameters which define a full binary solution it is important to only choose those which can be well determined from the given data. This resulted in several values being kept constant. Specifically, the temperatures of the two components are not well constrained due to the lack of eclipses and the presence of only two photometric colours. Additionally, third light was fixed to 15 percent for the BRITE red filter and 13.5 percent for the BRITE blue filter as determined in Sect. \ref{epslupb}\footnote{Though Sect. 3 quotes 15 percent due to the size of the error in the measurement, we use the actual number derived of 13.5 \%. We note though that the difference of 1.5 \% has no bearing on the values of the parameters derived. When allowed to vary they gave consistent parameters across the entire range of the uncertainty computed.}. While initially  we did attempt to fit third light contributions, however the fits were not sensitive enough to the value to provide useful information. 

\begin{figure*}
\begin{center}
    \includegraphics[scale=0.65]{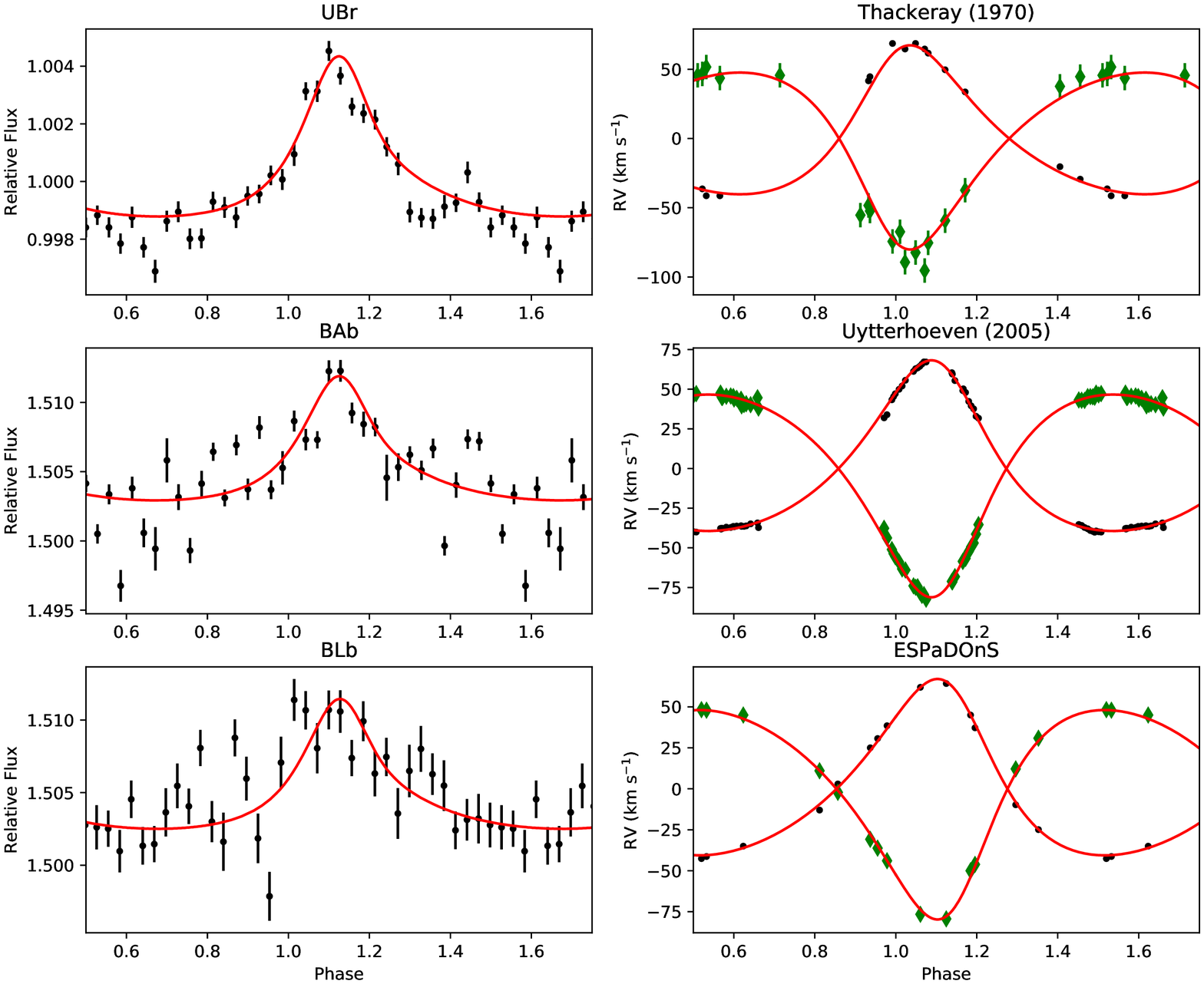}
    \caption{In the left column are phased light curves (black dots) from the three BRITE data sets, each binned in phase increments of $\approx 0.03$, to make the heartbeat phenomenon more apparent. In the right column are the primary (black dots) and secondary (green diamonds) for each of the three RV datasets. Overlaid on each of the 6 plots is the best fit simulation (red) to each dataset. $T_{0}$ is calculated with respect to the value in Table~\ref{tab:binary_params}. Since there is significant apsidal motion in this system, $T_{0}$ does not reflect periastron for all datasets.}\label{fig:bin_fit}

\end{center}
\end{figure*}

Even with the aforementioned exceptions the fit included a large parameter space comprised of 14 stellar and orbital parameters. We probed this space using 50 independent chains known as walkers. After a substantial burn-in to obtain the global minimum, the individual chains were allowed to walk for over 7000 iterations. The iterations were only allowed to stop when the parameter space was both well-sampled and all fit values passed the  Gelman-Rubin criterion for convergence \citep{gelman92}, i.e. when the in-chain variance is within 10\% of the variance between chains. The best fit for 3LCS can be seen in Fig.~\ref{fig:bin_fit} and the best-fit parameters of both fits are given in Tab. ~\ref{tab:binary_params} with 2$\sigma$ error bars. The only values not included are the passband luminosities for each light curve dataset, which were included solely for normalization purposes. While not fit directly, the masses and their corresponding errors we determined directly from Kepler's third law using period and semi-major axis distributions. The differences between our two solutions are minimal and all values are within errors. The only source of mild disagreement is with the inclination (and related parameters). The change in inclination, though small, leads to lower masses for both \el\ A components. Both fits are largely consistent with the parameters derived in \cite{2005A&A...440..249U}. We do measure a slightly larger apsidal motion and a slightly smaller period. Only the period does not agree within 2 $\sigma$ confidence likely meaning that the error implying that the error on this value is slightly underestimated by this work or \cite{2005A&A...440..249U}.

\subsection{Stellar parameters from evolutionary models}\label{sect:evol_model}

\begin{figure*}
\begin{center}
\includegraphics[width=18cm]{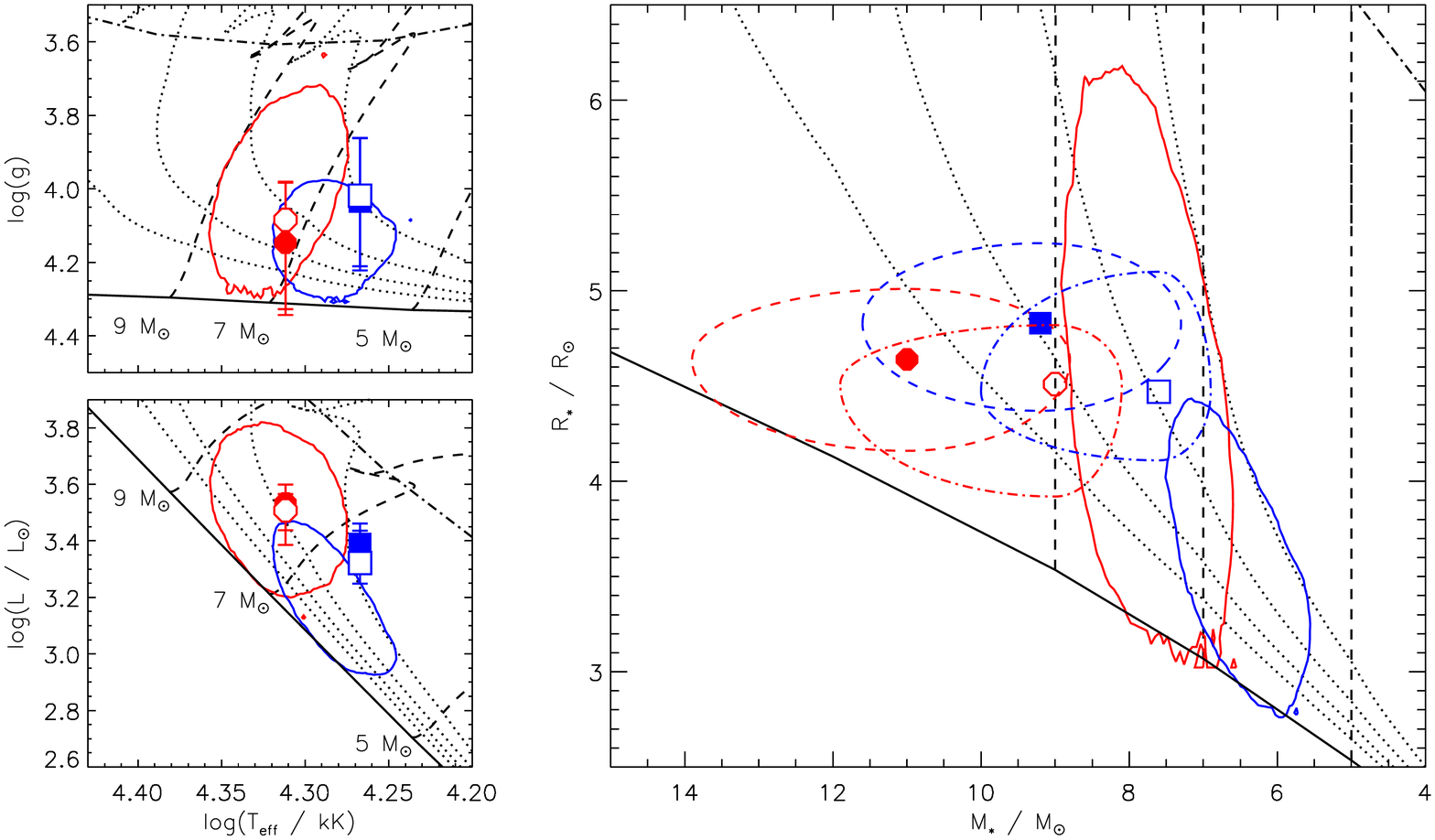}
\caption{Stellar parameters from spectroscopy and evolutionary models. In each panel the solid black line shows the ZAMS; the dot-dashed black line shows the TAMS; dashed black lines show evolutionary tracks; dotted black lines show isochrones, in intervals of $\log{(t/{\rm yr})}=0.2$, from 7.0 to 7.6. Solid contours show the 2$\sigma$ Monte Carlo point density (see text); Aa parameters are shown in red, Ab in blue. Parameters obtained from MCMC modeling of the full BRITE dataset are shown by the solid red circle (Aa) and solid blue square (Ab), with the dashed contours indicating the 2$\sigma$ uncertainties. Open symbols and dot-dashed contours are the parameters and uncertainties determined from the UBr dataset alone.}\label{fig:teff_logg_mass_rad}
\end{center}
\end{figure*}

As a point of comparison to the masses and radii determined via modeling the heartbeat variation, we also computed stellar parameters from evolutionary models. We adopted the grid of evolutionary tracks and isochrones calculated by \cite{ekstrom2012} for an initial rotation fraction of 0.4 of the critical rotation velocity. 

Since the luminosities of the two stars cannot be calculated directly from the system's $V$ magnitude and distance (since it is not obvious {\em a priori} what their individual contributions are), we instead started from their effective temperatures \teff~and surface gravities $\log{g}$. These quantities were determined by \cite{2019arXiv190202713S} via analysis of the ESPaDOnS spectra as $T_{\rm eff,Aa} = 20.5 \pm 0.5$~kK, $T_{\rm eff,Ab} = 18.5 \pm 0.5$~kK, $\log{g_{\rm Aa}} = 3.97 \pm 0.15$, and $\log{g_{\rm Ab}} = 4.13 \pm 0.15$. However, as noted in Sect.~\ref{epslupb}, these measurements assumed a 2-star model. We therefore adopted \teff~$21\pm1$ kK and $19\pm1$ kK for Aa and Ab (as in Sect.~\ref{epslupb}), and increased the uncertainty in $\log{g}$ by 0.05 to account for the element of uncertainty introduced by the B component.

Parameters were determined using a Monte Carlo algorithm, by populating the \teff-$\log{g}$ diagram with randomly generated points drawn from normal distributions in \teff~and $\log{g}$ corresponding to the value and uncertainty for each star's parameters, and then obtaining $M_*$, $\log{L}$, and stellar ages $t$ by linear interpolation between evolutionary tracks and isochrones, with radii $R_*/R_{\odot} = \sqrt{(L/L_\odot)/(T_{\rm eff}/T_{\rm eff,\odot})^4}$. Test points were accepted or rejected based on 3 criteria: 1) since the stars are presumably coeval, the ages of the test points must match within $\log{(t / {\rm yr})} = 0.1$; 2) the mass ratio $M_{\rm Aa}/M_{\rm Ab}$ of the test points must be within the uncertainty of the value determined from the RV curves; 3) the combined absolute magnitude $M_{\rm V}$ of the system must be within the uncertainty of the value determined from photometry. For the last two criteria, in order to ensure approximately Gaussian distributions, test points were compared to target values drawn from Gaussian distributions in $M_{\rm Aa}/M_{\rm Ab} = 1.19 \pm 0.01$ and $M_{\rm V} = -2.65 \pm 0.23$. 

The target value absolute $V$ magnitude is $M_{\rm V} = V - A_V - \mu$, where $A_{\rm V} = 0.04 \pm 0.04$ is the extinction \citep{2013MNRAS.429..398P}, and $\mu = 5\log{(d/{\rm pc})} - 5 = 5.9 \pm 0.2$ is the distance modulus. The distance $d = 156^{+19}_{-15}~{\rm pc}$ was obtained from the Hipparcos parallax $\pi = 6.4 \pm 0.7 {\rm \ mas}$ \citep{vanleeuwen2007}. To determine $M_{\rm V}$ for the test points, we first calculated bolometric magnitudes $M_{\rm bol} = M_{\rm bol,\odot} - 2.5\log{L/L_{\odot}}$, where $M_{\rm bol,\odot} = 4.74$. Bolometric corrections $BC$ were then applied according to \teff~and $\log{g}$, where we utilized the tabulated theoretical values calculated from non-LTE model atmospheres by \cite{lanzhubeny2007}, obtaining $BC = -1.95 \pm 0.05$ and $-1.78 \pm 0.05$ for the primary and secondary, respectively. The absolute magnitude of each component is then $M_{\rm V} = M_{\rm bol} - BC$, and the combined absolute magnitude is $M_{\rm V,tot} = -2.5 \log{(10^{-0.4M_{\rm V,Aa}} + 10^{-0.4M_{\rm V,Ab}}+ 10^{-0.4M_{\rm V,B}})}$. Since the atmospheric parameters of B are not well constrained, and it is therefore not obvious what $BC$ to use, rather than calculating $M_{\rm V,B}$ from $M_{\rm bol}$ we determined $M_{\rm V,B}$ using the flux ratio $f_{\rm B}/(f_{\rm B} + f_{\rm A}) = 0.15 \pm 0.1$ found above in Sect.\ \ref{epslupb}. $f_{\rm B}$ was drawn from a Gaussian distribution corresponding to the flux ratio, and $M_{\rm V,B}$ calculated from $M_{\rm V,A}$ under the assumption that the flux ratio is relatively flat between the blue and red BRITE filters.

The algorithm is terminated when $10^4$ points have been accepted, at which point stellar parameters are determined from the peaks of posterior Probability Density Functions (PDFs). The 2$\sigma$ density contours of the grids are shown in Fig.\ \ref{fig:teff_logg_mass_rad}. The algorithm finds $M_{\rm V,Aa} = -2.0 \pm 0.3$ mag, $M_{\rm V,Ab} = -1.4 \pm 0.3$ mag, $M_{V,B} = -0.4 \pm 1.1$ mag, $M_{\rm bol, Aa} = -4.1 \pm 0.3$ mag, $M_{\rm bol, Ab} = -3.3 \pm 0.3$ mag, $\log{L_{\rm Aa}}/L_{\odot} = 3.5 \pm 0.1$, $\log{L_{\rm Ab}/L_{\odot}} = 3.2 \pm 0.1$, $M_{\rm Aa} = 7.6 \pm 0.4~ M_\odot$, $M_{\rm Ab} = 6.4 \pm 0.4~ M_\odot$, $R_{\rm Aa} = 4.3 \pm 0.6~ R_\odot$, $R_{\rm Ab} = 3.5 \pm 0.4~ R_\odot$, and $\log{(t/{\rm yr})} = 7.45 \pm 0.15$. 

The three panels of Fig.\ \ref{fig:teff_logg_mass_rad} also show the values and 2$\sigma$ uncertainties obtained from MCMC modeling of the photometric and radial velocity variations for comparison, using both the full BRITE dataset and the UBr dataset only. The surface gravities derived from MCMC modeling overlap with those determined spectroscopically. Luminosities are also approximately consistent with the values via evolutionary models, although the Ab component's MCMC luminosity is somewhat higher. The Aa component's radius is compatible with the MCMC value; however, the radius of the Ab component is larger than the evolutionary model value. MCMC masses are systematically higher than evolutionary model masses. Notably, the masses and radii obtained from the UBr dataset alone overlap with the evolutionary model parameters obtained for both stars. 

\subsection{Constraints from interferometry}\label{sect:interferometry}

We fit the interferometric observables of \el~A with a binary model. The individual diameters were forced to the expected apparent size of 0.25~mas and 0.23~mas. However, the exact values used have no significant impact on the results because these diameters are unresolved even by the longest baseline of VLTI. 

\begin{table}
\centering
\caption[Interferometric observations]{Interferometric observations from PIONIER/VLTI. The astrometric error ellipse is given by its semi-major ($e_{max}$) and semi-minor ($e_{min}$) axes, and the position angle of its major axis. PA and Sep are the position angle and separation of the B with respect to the A component brightest in $H$-band. PA is measured eastwards from North.}
\begin{tabular}{llllll}
\hline
HJD      & Sep & PA & $e_{min}$ & $e_{max}$ & $PA_{max}$ \\
$-2450000$ & (mas) & ($^{\circ}$) & (mas) & (mas) & ($^{\circ}$) \\
\hline
6818.024 & 1.02 & 41.4 & 0.24 & 0.17 & 92 \\
6818.155 & 1.03 & 39.2 & 0.21 & 0.15 & 168 \\
6819.080 & 0.92 &  83.1 & 0.32 & 0.14 & 131 \\
\hline
\hline
\end{tabular}
\label{vlti_tab}
\end{table}

The remaining free parameters are the two coordinates of the apparent separation vector (East and North, in units of mas) and the flux ratio in $H$-band between the secondary and the primary. The best fit is obtained for a flux ratio ($\frac{A_{B}}{A_{A}}$) of $0.55\pm0.09$, considered constant over the $H$-band. Note that the flux ratio is partially degenerate with respect to the angular separation because the latter is just barely larger than the angular resolution of the observations. Inferred positions are summarized in Table~\ref{vlti_tab}.

\begin{figure}
\includegraphics[scale=0.7,trim={0 0 2.75cm 0},clip]{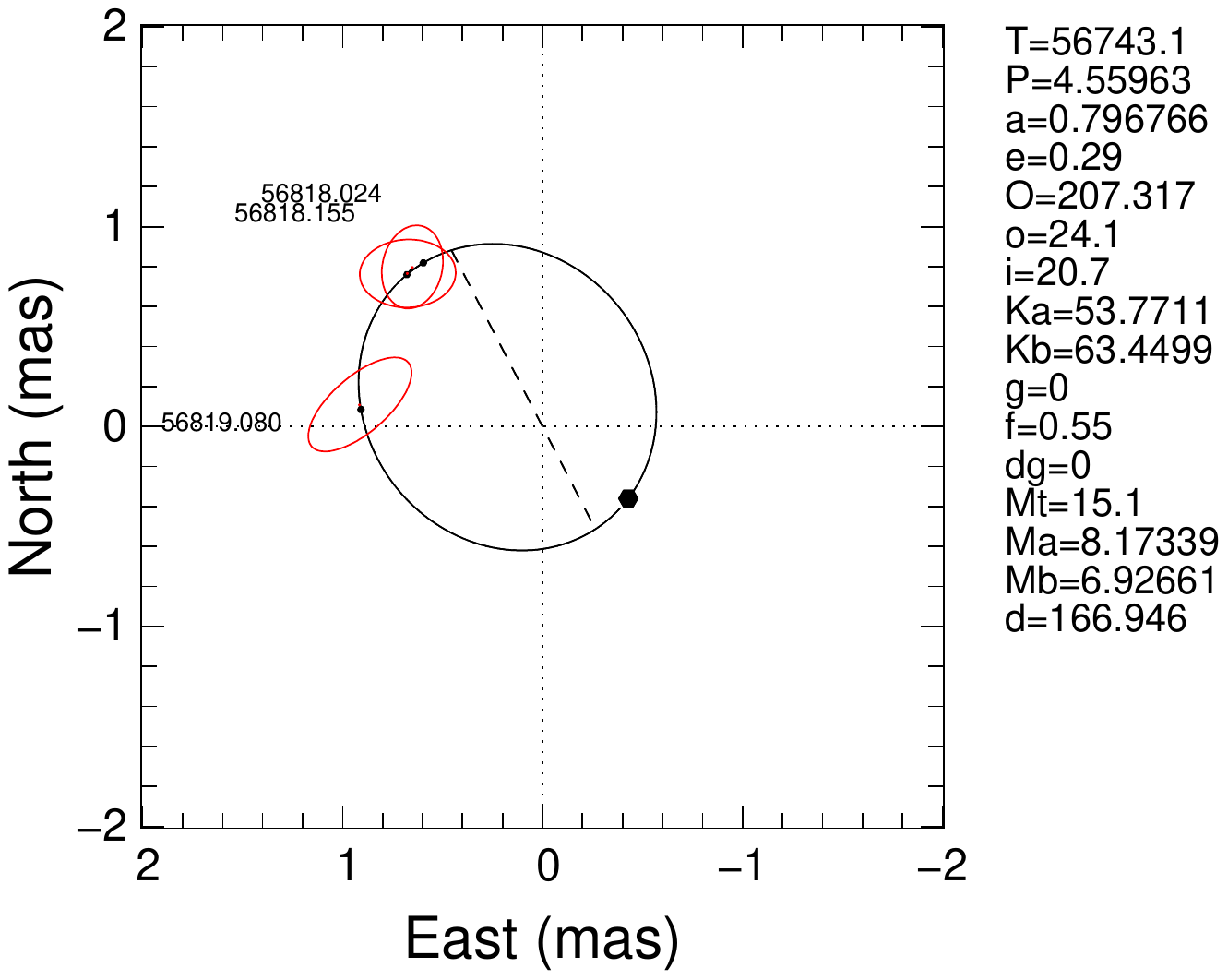}
\caption{Motion of the secondary around the primary as spatially resolved by our PIONIER observations. The orbit is given by the solid line with the label on each error ellipse representing the MJD of the observation. The periastron of the secondary is represented by a filled symbol and the line of nodes by a dashed line.\label{fig:orbit_interfero}}
\end{figure}

The interferometric observations are too sparse to independently recover all the orbital elements without additional constraints. Therefore, we impose all previously constrained parameters ($M_{1,2}$, $P_{orb}$, $T_{0}$, $e$, $\omega$, $i$) and adjust only the position angle of the ascending node ($\Omega$) and the parallax of the system (distance). The latter is given by the ratio between the apparent size of the orbit constrained by interferometry, and the physical size of the orbit imposed by the total mass and the period. The best fit apparent orbit is shown in Figure~\ref{fig:orbit_interfero}. The inferred parallax of 5.99~mas(no error estimate available) is within 0.5 $\sigma$ from the $6.37\pm0.7$~mas parallax from Hipparcos. 

To compare the $H$-band flux ratios with the luminosities determined above, we converted the $V$-band absolute magnitudes determined in Section \ref{sect:evol_model} ($M_{V} = -1.6$ and $-2.1$ for the secondary and primary, respectively) to $H$-band absolute magnitudes using the empirical main-sequence colour/effective temperature table published by \cite{2013ApJS..208....9P}. The expected $H$-band absolute magnitudes are, for the secondary and primary,  $-1.1$ and $-1.5$, yielding an $H$-band flux ratio of 0.65. The 10\% difference between this value and the interferometric flux ratio can easily be accounted for by the uncertainties in the luminosity, as within this range the $H$-band flux ratio can vary between 0.43 and 0.98. 

\subsection{Comparison of Results}

It is clear that while there is some overlap, the  evolutionary and binary models give different values for the main stellar parameters, namely the masses and radii of the two components. While interferometric measurements should help us to identify which model is more accurate, the number of such observations is simply insufficient. We are thus left with the unenviable task of trying to reconcile these differences. As any real discrepancy would be extremely important to our knowledge of stellar evolution and magnetic fields, it must be considered carefully. 

The cause of the observed discrepancy in mass is largely due to the value of the inclination. While the binary fit for the 3LCS and 1LCS solutions prefer a value of around $19^{\circ}$ and $20^{\circ}$ respectively, to achieve the masses and radii preferred by the evolutionary models would require an inclination closer to $22^{\circ}$. Therefore, we adjusted the inclination, and concurrently the semi-major axis as these two parameters are highly degenerate,  keeping all other parameters fixed and found that we could find a reasonable by-eye fit to both photometric and RV datasets at this inclination. This is confirmed when examining the $\chi^{2}$ value corresponding to , as this value is only slightly worse than for the best-fit model.  However, the MCMC sampling considers this difference significant as even when constraining the inclination to a small parameter space around  $22^{\circ}$ during fitting it will try to converge to values outside of this range. 

Our ability to achieve a reasonable fit at $22^{\circ}$ would seem to imply that the discrepancies noted between the evolutionary and binary methods are not significant. However, the fact that these two methods converge to different values must be explored further.  One possible cause can be seen in the phased UBr light curve shown in  Fig.~\ref{fig:bin_fit}.  Close inspection shows what appears to be an isolated brightness maximum, around phase 1.45. Such coherent variability in phase would not be unexpected in heartbeats due to the nature of tidally excited oscillations (see Sect.~\ref{sect:TEOs}).  Whether this variability is a sign of such oscillations or simply correlated noise, it could have an effect on the binary fit.  Unfortunately, while there are some low-level signals  present in the phased light curve, their removal does not affect the residuals in any noticeable way.  Since there is no way to easily remove this variability, we instead focused on removing the largest amplitude non-binary variability around phase 1.45. This could be artificially enhancing the width of the heartbeat shape which will affect the inclination. Therefore, we cut out a region 0.1 in phase around this peak in all three light curves and tried the fitting procedure outlined in Sect.~\ref{sect:model} on the adjusted data. Despite these efforts the fit converged to roughly the same values.  While we believe that this variability is still the most likely source of our discrepancy we have no way of quantifying this hypothesis with the data currently available. In Sect.~\ref{sect:mag},  we will explore the effect that magnetism could have had on the system's evolution and whether such a discrepancy might legitimately be expected.

\section{Search For Additional Variability}

While binarity is the strongest source of photometric variability, it is clear from Fig.~\ref{full_ft} that there are likely other signals present. To check the validity of these signals we first removed the binary variability.  For the UBr data this is done by subtracting off a PHOEBE simulation (see Sect.~ \ref{sect:model}).  As the SMEI transmission curve is not available within the PHOEBE framework, we instead created a template using the binned, phase-folded SMEI data  which was then subtracted from the original data. We then determined a point-error-weighted frequency spectrum  of the subtracted light curves shown in Fig.~\ref{fig:ft}.  We are now able to explore these signals and their likely causes.

\subsection{Frequency Determination}

\begin{figure*}
\includegraphics[scale=0.8]{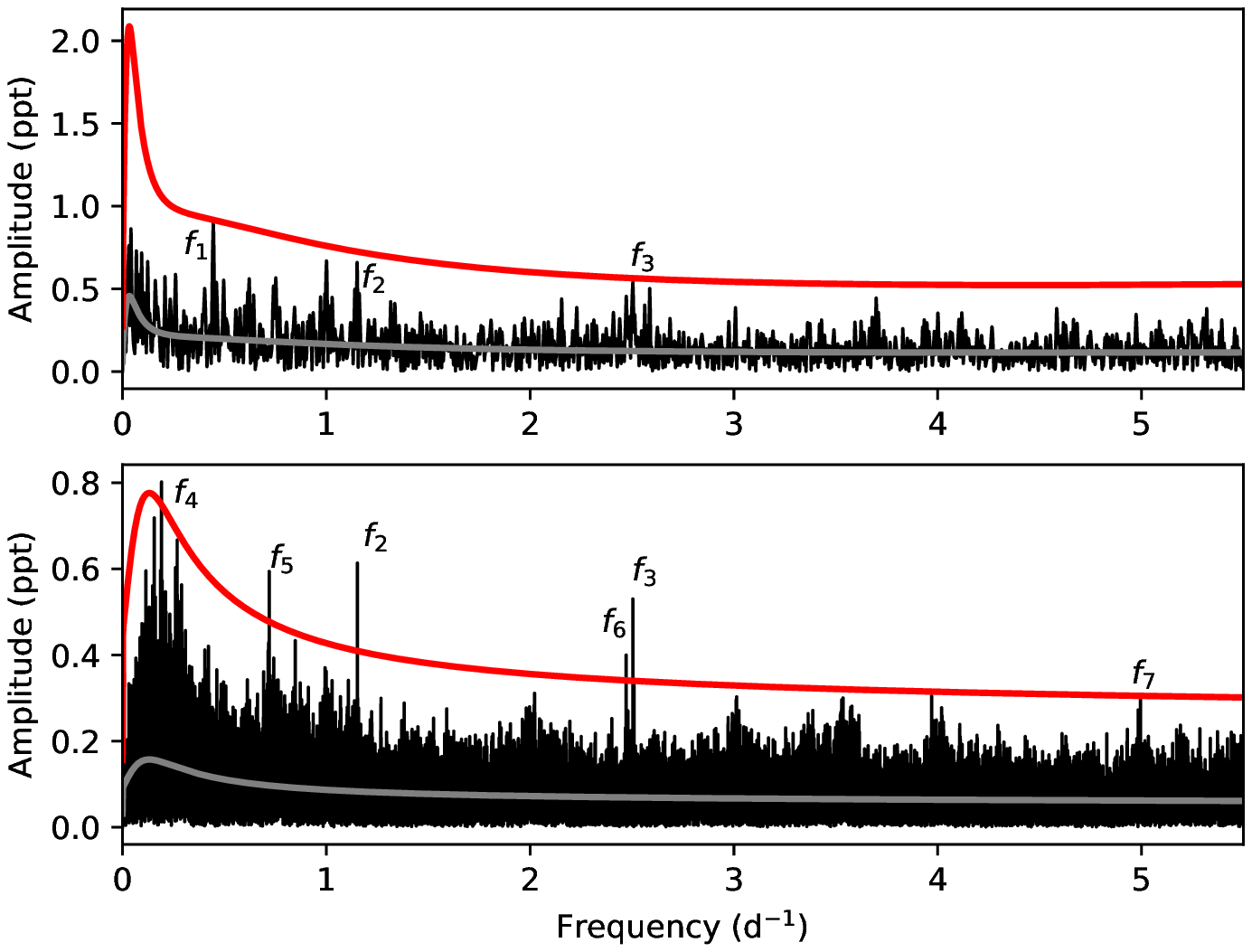}
\caption{Discrete point-error weighted FS of data from the BRITE UBr satellite (top) and SMEI satellite (bottom) with binary variability removed.  In each plot there are lines indicating the noise floor (grey) and the significance threshold (red).}\label{fig:ft}
\end{figure*}

Frequency determination first required selection of  a significance criterion. For this we chose  the False Alarm Probability ($FAP$)  as outlined by \cite{fap_ref}. This denotes the probability according to Gaussian statistics that a peak of a given height will occur due to noise. In our case, we chose the significance threshold such that in our data sample we expect less than one such peak to be present. For the UBr dataset used in our frequency analysis this equates to 0.016\%, while for the SMEI data this is 0.00085\%.  The amplitude of the peak corresponding to this $FAP$ is not constant across all frequencies because the noise floor is not constant. This is especially true at frequencies at or below 1 d$^{-1}$  which are of interest for $\varepsilon$ Lupi.  We calculate the noise floor by fitting  the power density of the FS to the following function:

\begin{equation}
PD = \frac{A}{1+(\tau f)^{\gamma}} + c\rm{,} 
\end{equation}    
where $c$ is the constant white noise, $A$ is the amplitude, $\tau$ is the characteristic timescale associated with the signal, $f$ is the frequency, and $\gamma$ is the power index \citep{noise_floor}.  The FT along with its noise floor and significance threshold is shown in Fig.~\ref{fig:ft}.

The individual frequencies were determined using the standard pre-whitening procedure. This is an iterative fitting process where the location of each peak is determined and then removed by fitting the data using a sinusoidal fit corresponding to that frequency, its phase and amplitude. This process is then repeated refitting the sum of the sinusoids of all the determined peaks, allowing all fitted parameters to vary, until no peak remains above the significance threshold.  This resulted in 7 unique significant frequencies which are given in Tab.~\ref{table:freq} and Tab.~\ref{table:freq_smei}.  Of these, two frequencies $f_{4}$ and $f_{7}$ are likely instrumental. $f_{4}$ is an integer number of the yearly frequency from the orbital period, while $f_{7}$ is an integer number of yearly aliases from $5~ \rm c~d^{-1}$.   Of the remaining,  $f_{2}$,  $f_{3}$, and $f_{6}$ appear in both the UBr and SMEI datasets, though $f_{6}$ is just below the detection threshold in UBr.   Finally, there are two frequencies $f_{1}$ and $f_5$ which only appear in UBr, and SMEI respectively. 

\begin{table*}
\begin{center}
\caption{Significant Frequencies from UBr data of \el. Phase represents the shifted zero point of the fitted sinusoid for each frequency relative to an arbitrary fixed point near the start of observations. FAP threshold is 0.016\%. Errors were calculated from 10000 MCMC iterations. }\label{table:freq}
\begin{tabular}{c  c c c c}
\hline
Number &  Frequency (d$^{-1}$) & Amplitude (ppt) & Phase & $FAP$ (\%) \\
\hline
$f_{1}$ & $0.44500 \pm 0.00098$ & $0.89^{0.21}_{-0.21}$ & $0.60^{0.39}_{-0.39}$ &  $0.0026$ \\
$f_{2}$ & $1.151 \pm 0.012 $ & $0.67^{0.22}_{-0.45}$ & $0.0^{0.27}_{-0.09}$ & 0.0021\\
$f_{3}$ &  $2.504 \pm 0.17$ & $0.53^{0.22}_{-0.34}$ & $0.20^{0.31}_{-0.16}$ & 0.011\\
\hline
\hline
\end{tabular}
\end{center}
\end{table*}

\begin{table*}
\begin{center}
\caption{Significant Frequencies from SMEI data of \el. Phase represents the shifted zero point of the fitted sinusoid for each frequency relative to an arbitrary fixed point near the start of observations. FAP threshold is  0.00085\%. Errors were calculated from 10000 MCMC iterations.}\label{table:freq_smei}
\begin{tabular}{c  c c c c}
\hline
Number &  Frequency (d$^{-1}$) & Amplitude (ppt) & Phase & $FAP$  (\%) \\
\hline 
$f_{4}$& $0.19172 \pm 4\times10^{-5}$ & $0.81 \pm 0.16 $ & $0.70 \pm 0.032 $ & $4.10\times10^{-6}$ \\
$f_{2}$ &   $1.15278 \pm 5.3\times10^{-5}$ & $0.63 \pm 0.16$ & $0.503 \pm 0.042$ & $< 1 \times 10^{-10}$ \\
$f_{5}$ &   $0.71985 \pm 6\times10^{-5}$ & $0.60 \pm 0.16$ & $0.895 \pm 0.044$ &  $7.3\times10^{-10}$ \\
$f_{3}$ &   $2.50421 \pm 6\times10^{-5}$ & $0.56 \pm 0.16$ & $0.85 \pm 0.05$ &  $< 1 \times 10^{-10}$ \\
$f_{6}$&   $2.47154 \pm 8\times10^{-5}$ & $0.45 \pm 0.16 $ & $0.93  \pm 0.06$ &  $< 1 \times 10^{-10}$\\
$f_{7}$ &   $4.9945 \pm 2\times10^{-4}$ & $0.33 \pm 0.16$ & $0.5 \pm 0.1$ & 0.00017 \\

\hline
\hline
\end{tabular}
\end{center}
\end{table*}

One common feature in heartbeat systems is the existence of tidally excited oscillations (TEOs) which often appear at integer factors of the orbital frequency. However, this phenomenon is notably absent in \el. While there is one frequency which is close, $f_{1}$, the error bars are narrow enough to discount it from being an exact multiple of the orbit.  Moreover, this frequency is conspicuously absent in the SMEI data. The lack of stability is important as tidally excited oscillations are constant in phase, frequency and amplitude on timescales of years \citep{oleary:14,guo16}.  The ramifications of this are discussed in more detail in Sect.~\ref{sect:TEOs}. \cite{2005A&A...440..249U} note a similar period to $f_{1}$ in the equivalent widths of the primary, so pseudo-synchronous rotation is a distinct possibility. 

For all other non-instrumental frequencies,  it is reasonable to suspect that they can be attributed to pulsation. To test this idea, an evaluation of the pulsation constant $Q\,=\,P\,\sqrt{\overline{\rho}/\overline{\rho}_{\odot}}$ is useful. While we quote two models with decidedly different masses, the effect on $Q$ is negligible so we will focus on the stellar parameters for the 1LC model listed in Table ~\ref{tab:binary_params}. We pair these with the frequencies listed in Table \ref{table:freq_smei}, ignoring those of instrumental origin. We derive $0.126$\,d\,$<Q<0.400$\,d for the primary, and $0.116$\,d\,$<Q<0.368$\,d for the secondary. In the region of the HR Diagram where the components of \el ~are located, two classes of pulsating stars can be found, the $\beta$~Cephei stars and the slowly pulsating B (SPB) stars. The former pulsate in low-order pressure and mixed modes, whereas the latter oscillate in gravity modes of longer period. This pulsational behaviour translates into $Q<0.04$\,d for the $\beta$~Cephei stars \citep{2005ApJS..158..193S}, whereas the $Q$ values for SPB stars are considerably larger than that. For this reason we conclude that the periodic signals present in the light curves suggest SPB-type gravity modes. This result is in strong contrast to those of \cite{2005A&A...440..249U}. However, their results are based largely on the presence of a frequency at $10.36~\rm{d}^{-1}$ of which there is no evidence in our data. Interestingly, we both show evidence of a low amplitude peak at $6.46~\rm{d}^{-1}$, but since it is not significant in this work or that of \cite{2005A&A...440..249U}, it is impossible to speculate further.   


\subsection{Tidally Excited Oscillations} \label{sect:TEOs}

TEOs are a common feature of heartbeat systems, and are of particular interest in \el .  Recent work \citep{fuller:15,stello:16,cantiello:16,lecoanet:17} has shown that gravity waves cannot propagate in the presence of strong magnetic fields, and hence standing gravity modes do not exist in strongly magnetized stars. Hence, TEOs due to gravity modes may be suppressed in $\varepsilon$~Lupi; this is what we seek to determine here. To do this, we compute expected TEO amplitudes in the absence of a magnetic field to see if we should have observed TEOs in an equivalent non-magnetic binary. We model TEOs in $\varepsilon$~Lupi in the same manner as \cite{pablo:17} did for $\iota$~Orionis.  We first construct stellar models that are in the approximate range of the spectroscopic and light curve modeling results from above. Here we examine a model with primary mass $M=7.9 \, M_\odot$ and $R=5.3 \, R_\odot$. We assume a rotation period of $4$ days, with a rotation axis aligned with the orbital axis. As the stellar and spectroscopic models have some clear discrepancies we have chosen a model which is optimistic in terms of TEOs, as higher stellar masses or smaller radii as preferred by the light curve modeling (see Table~\ref{tab:binary_params} predict TEO amplitudes lower by a factor of $\sim \! 2$. After creating stellar models, we computed their gravity modes using GYRE \citep{townsend:13} and computed their tidally forced amplitudes as described by \cite{fuller17}. We only computed TEOs produced in the primary star; modes of the secondary star are expected to have smaller amplitudes due to its smaller radius and smaller contribution to the luminosity of the system.

\begin{figure}
\includegraphics[scale=0.5]{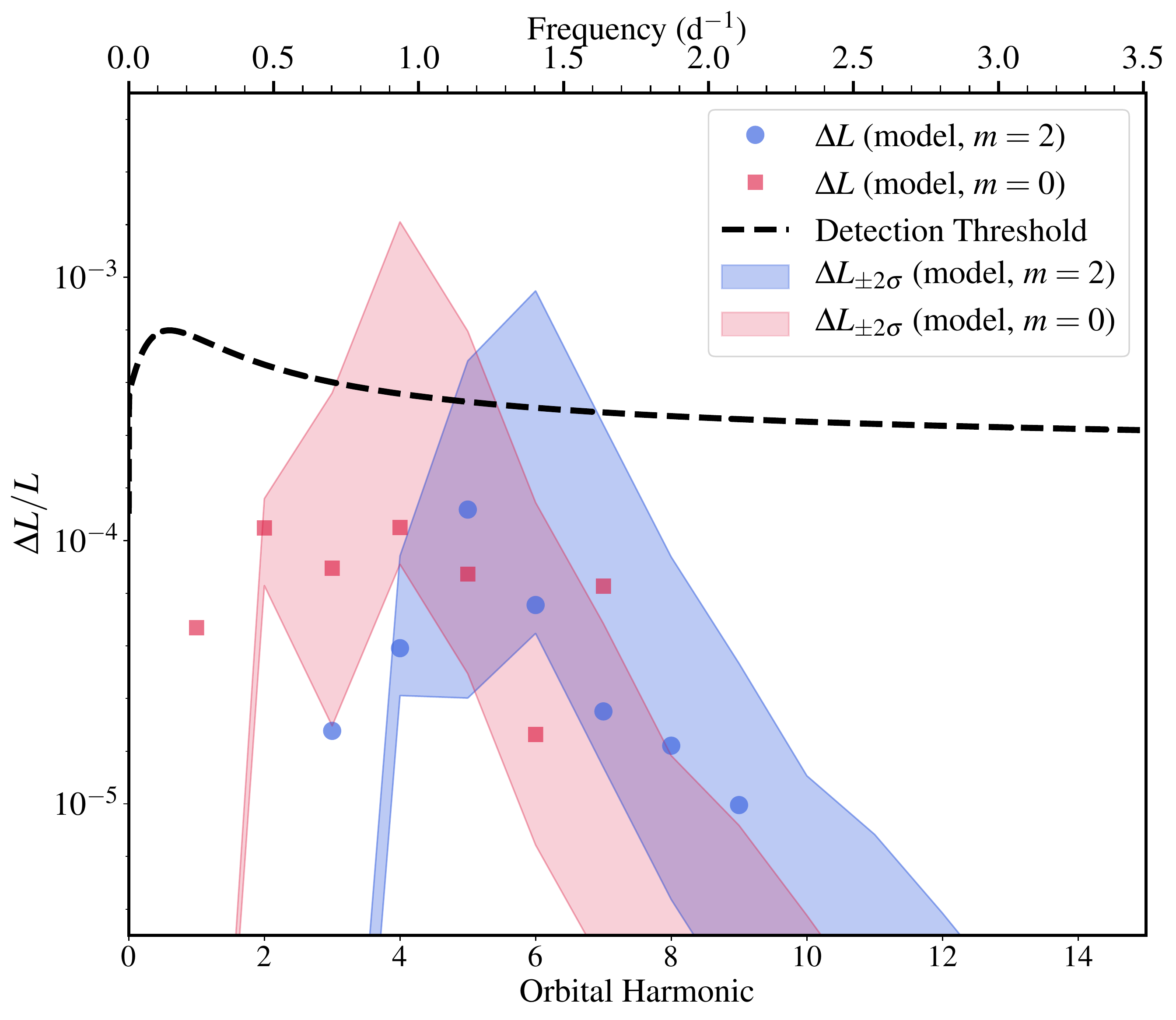}
\caption{Tidal model of $\varepsilon$ Lupi showing predicted amplitudes of tidally excited oscillations as a function of frequency. Blue circles and red squares are $m=2$ and $m=0$ modes, respectively. Shaded areas of the corresponding colour denote where we expect 95 \% of tidally excited oscillations to occur. We only plot the contribution from the primary star, as the secondary is expected to contribute at lower amplitudes. Comparison with Figure \ref{fig:ft} shows that we expect most tidally excited oscillations to lie below our detection threshold.}\label{fig:EpsLupTide}
\end{figure}

Figure \ref{fig:EpsLupTide} shows our predicted TEO amplitudes as a function of frequency for $\varepsilon$ Lupi. The shaded region shows where we expect $95 \%$ of TEOs to exist using the theory of \cite{fuller17}, which accounts for the probability of resonantly excited modes.This theory breaks down at frequencies below 2 and 4 times the orbital frequency for $m=0$ and $m=2$ modes, respectively, where gravity modes become traveling waves. However, we expect low TEO amplitudes in this regime anyway.  Figure \ref{fig:EpsLupTide} shows that TEOs are expected to have amplitudes  $\Delta L/L < 10^{-3}$ at all frequencies, and that most TEOs should have amplitudes $\Delta L/L \lesssim 4 \times 10^{-4}$  which is below our detectability threshold. Hence, we do not expect to detect TEOs in \el\ with the current dataset. However, if the detection threshold can be decreased to $\Delta L/L \sim 10^{-4}$, we might expect to detect TEOs with frequencies $f \lesssim 1.8 \, {\rm d}^{-1}$. 

As demonstrated by e.g. \cite{pablo:17}, we can also use our TEO models to estimate the energy dissipation rate due to TEOs. For \el, we estimate a circularization timescale of $\sim 2 \times 10^8 \, {\rm yr}$, longer than the $\sim 1.8 \times 10^7 \, {\rm yr}$ age of the system  based on the stellar masses and radii. However, we estimate a pseudosynchronization time scale of $\sim 4~{\rm Myr}$. These estimates are consistent with the measured eccentricity of $\varepsilon$~Lupi. They also suggest that the spins of the stellar components may have been tidally pseudosynchronized, although magnetic interactions could also potentially contribute to this process. This will be discussed in the following section.

It is unlikely that we would be able to detect any potential TEOs as they are predicted to be below our significance threshold. Therefore, we are unable to provide any insight into the effect of the magnetic field on TEOs.  However, we do detect one anomalous frequency $f_{1}$ near the 2nd orbital harmonic. It is possible this is an $m=0$ tidally excited g-mode, a tidally excited r-mode (which are not included in our calculations), a residual from an imperfect lightcurve model, or an unrelated stellar pulsation. More data are needed to reach a firm conclusion.   

\section{Understanding interactions within the system} \label{sect:mag}
$\varepsilon$ Lupi is the only short-period doubly-magnetic massive binary currently known. Therefore, two dominant types of interactions could be in play: tides (as has been discussed in the previous section) and direct electromagnetic interactions between detected stellar magnetic fields. In this work, we will not discuss the potential interactions between the winds and the magnetospheres of the two stars. 

The key questions to answer for tidal and electromagnetic interactions are: How do they compete? 
Do these mechanisms allow us to understand the observed orbital eccentricity and inclination, the rotation state of the components, and the observed magnetic dipole orientations?\\

To answer these questions, we have to compute the order of magnitude of the strength of each type of interaction. Very few studies have simultaneously considered tidal and magnetic interaction forces and torques \citep[][]{King1990,Campbell1997,Strugarek2017}; moreover, they were focused on the specific case of cataclysmic variables or close star-planet systems.

First, we assume that each stellar magnetic field can be modeled outside the star as a dipolar field, a reasonable assumption for supposed fossil fields \citep[e.g.][]{BS2004,BN2006,DM2010}, with its radial and colatitudinal components and amplitude respectively expressed as
\begin{eqnarray}
&&B_{i;r}\left(r,\theta\right)=\frac{\mu_0}{4\pi}{\mathcal M}_{i}\frac{2\cos\theta}{r^3},
B_{i;\theta}\left(r,\theta\right)=\frac{\mu_0}{4\pi}{\mathcal M}_{i}\frac{\sin\theta}{r^3}\nonumber\\
&&B_{i}\left(r,\theta\right)=\frac{\mu_0}{4\pi}{\mathcal M}_{i}\frac{1}{r^3}\left(3\cos^2\theta+1\right)^{1/2},
\end{eqnarray}
where $r$ is the radius, $\theta$ the co-latitude, $\mu_0$ the magnetic permeability of  the vacuum (we work here in SI) and ${\mathcal M}_i$ the magnetic dipolar moment of the field of the $i\equiv\left\{1,2\right\}$th stellar component. This allows us to express ${\mathcal M}_i$ as a function of the observed polar magnetic field
\begin{equation}
{\mathcal M}_{i}=\frac{2\pi R_{i}^{3}}{\mu_{0}}B_{i}^{\rm p}\quad\hbox{with}\quad B_{i}^{\rm p}=B_{i}\left(R_{i},0\right),
\end{equation}
where $R_{i}$ is the radius of the $i$th star. The interaction energy between two magnetic dipoles placed at a distance $r_{12}$ from each other is given by
\begin{eqnarray}
{\varepsilon}_{1-2}^{\rm mag}&=&
\frac{\mu_{0}}{4\pi}\frac{\left[{\vec{\mathcal M}}_{1}\cdot{\vec{\mathcal M}}_{2}-3\left({\vec{\mathcal M}}_{1}\cdot{\vec e}_{12}\right)\left({\vec{\mathcal M}}_{2}\cdot{\vec e}_{12}\right)\right]}{r_{12}^3}\nonumber\\
&=&\frac{\mu_0}{4\pi}\frac{\left[{\mathcal M}_{1}{\mathcal M}_{2}\cos\left(\Theta_{12}\right)-3{\mathcal M}_{1}\cos\left(\Theta_1\right){\mathcal M}_{2}\cos\left(\Theta_2\right)\right]}{r_{12}^3},\nonumber\\
&\equiv&\frac{\mu_0}{4\pi}\frac{{\mathcal M}_{1}{\mathcal M}_{2}}{a^3},
\label{eq:emag}
\end{eqnarray}
where ${\vec e}_{12}=\frac{{\vec r}_{12}}{r_{12}}$,  $\Theta_{i}$ (with $i\equiv\left\{1,2\right\}$) is the obliquity of $\vec{\mathcal M}_{i}$ relative to ${\vec e}_{12}$, $\Theta_{12}=\Theta_{2}-\Theta_{1}$ is their relative obliquity, and we have introduced the orbital semi-major axis $a$ to provide an order of magnitude for $r_{12}$. This is an approximation for an eccentric orbit such as that of $\varepsilon$ Lupi. However, this is sufficient to provide orders of magnitude. The corresponding force is given by \citep[see also][]{King1990}
\begin{equation}
{F}_{1-2}^{\rm mag}=\vert\!\!\vert-{\vec\nabla}{\varepsilon}_{1-2}^{\rm mag}\,\vert\!\!\vert\equiv\frac{\mu_0}{4\pi}\frac{{\mathcal M}_{1}{\mathcal M}_{2}}{a^4}=\frac{\pi}{\mu_0}B_{1}^{\rm p}B_{2}^{\rm p}\frac{R_{1}^{3}R_{2}^{3}}{a^4}.
\end{equation}
On the other hand, the strength of the tidal force is given by \citep[e.g.][]{Murray2000}
\begin{equation}
F_{1-2}^{\rm tide}\equiv\frac{G M_1 M_2}{a^2}\frac{R_1}{a}.
\end{equation}

Note that for each of the interaction forces, we focus on the dependence on stellar quantities and on the distance between stars. The angular dependences are filtered out to simplify the problem and to derive order of magnitude estimates, which are sufficient for our purpose.

The relative strength of the electromagnetic and tidal forces is thus given by
\begin{equation}
\label{Rmagtide}
{\mathcal R}_{\rm{mag/tide}}\equiv\frac{\pi}{\mu_0 G}\frac{1}{M_1 M_2}B_{1}^{\rm p}B_{2}^{\rm p}\frac{R_1^2 R_2^3}{a},
\end{equation}
where we express the orbital semi-major axis ($a$) as a function of the orbital period ($P_{\rm orb}$) using  Kepler's third law
\begin{equation}
a=\left[G\left(M_1+M_2\right)\left(\frac{P_{\rm orb}}{2\pi}\right)^2\right]^{1/3}.
\end{equation}
Using the values for polar magnetic fields provided by \cite{2015MNRAS.454L...1S} (i.e. $B_1=600$G and $B_2=900$G), we obtain ${\mathcal R}_{\rm{mag/tide}}\approx 5.29 \times 10^{-12}$. Therefore, the electromagnetic interaction force is very small compared to the tidal force. We conclude that the orbital evolution of the system and the rotational evolution of its components should be completely driven by tides \footnote{As pointed out in the previous section, stellar magnetic fields can modify TEOs that would indirectly impact corresponding synchronization, alignment and circularization times \citep[see also][]{Wei2016,LinOgilvie2018}.}. 

It is also constructive to compare magnetic and tidal energies, rather than magnetic/tidal forces. The characteristic energy of the tidal interaction with the primary star (i.e., the potential energy in its equilibrium tidal bulge) is 
\begin{equation}
\varepsilon^{\rm tide}_{1-2} = \frac{G M_2^2 R_1^5}{a^6} \, .
\end{equation}
The tidal interaction energy of the second star is the same with $1-2$ subcripts reversed, and is the same order of magnitude for $\varepsilon$ Lupi. We then find an interaction energy ratio
\begin{equation}
\label{Emagtide}
\frac{\varepsilon^{\rm mag}_{1-2}}{\varepsilon^{\rm tide}_{1-2}} = \frac{\pi B_1 B_2 R_2^3 a^3}{\mu_0 G M_2^2 R_1^2} \, .
\end{equation}
For $\varepsilon$ Lupi, $M_1 \sim M_2$, and equation \ref{Emagtide} is larger by a factor $\sim a^4/R_1^4$ compared to equation \ref{Rmagtide}. However, equation \ref{Emagtide} evaluates to $ \sim 4 \times 10^{-6}$, and so magnetic effects are still dominated by tidal effects.

The sole direct impact of the electromagnetic interaction force would therefore be on obliquities of the magnetic axes. In this framework, the lowest stable energy state is the  horizontal aligned magnetic-spin configuration $\left(\rightarrow\quad\rightarrow\right)$  (the arrows being here the magnetic spins) for which $\Theta_{1}=\Theta_{2}=\Theta_{12}=0$ in Eq. \ref{eq:emag} and ${\varepsilon}_{1-2}^{\rm mag}=-\mu_0/4\pi\,\left(2{\mathcal M}_{1}{\mathcal M}_{2}\right)/r_{12}^3$. However, as pointed out by \cite{2015MNRAS.454L...1S}, the obliquity of the field with respect to the rotation axis in each star is assumed to be small. Therefore, if tides have already contributed to align the  rotation spins of the stars and the orbital spin, the magnetic-field directions should be parallel. In this configuration, the lowest-energy stable magnetic configuration due to the magnetic dipole-dipole interaction force is vertical anti-aligned magnetic fields $\left(\uparrow\,\,\downarrow\right)$,  where $\Theta_1=\pi/2$, $\Theta_2=-\pi/2$, $\Theta_{12}=-\pi$, and ${\varepsilon}_{1-2}^{\rm mag}=-\mu_0/4\pi\,\left({\mathcal M}_{1}{\mathcal M}_{2}\right)/r_{12}^3$ (see Eq. \ref{eq:emag})
, which corresponds well to $\varepsilon$ Lupi observations.

\section{Discussion \& Future Work}

In this work we have provided an in-depth analysis of the \el\ system. For the first time, we have found spectroscopic evidence of a tertiary component (\el\ B) which has heretofore escaped notice due to its large rotational velocity. We were also able to determine empirically the fundamental parameters of both Aa and Ab despite both a low system inclination and modest eccentricity. This is a testament to just how valuable heartbeat systems can be in the understanding of massive stars, where fundamental parameters are difficult to come by. 

The modest binary variability, however, is at the limit of what is possible with the BRITE satellites and is likely the cause for the discrepancy in parameter values seen when compared to stellar evolution models. This is further confirmed by our analysis of magnetic interactions in this system. As these are small compared to tidal effects, the evolutionary history of this system should be similar to other non-magnetic binary systems. 

The one caveat to this statement is the presence, or absence, of tidally excited gravity modes. Recent work has shown that gravity modes are suppressed in stars with sufficiently strong magnetic fields \citep{fuller:15,stello:16,cantiello:16,lecoanet:17}. Similar to the $5 \, M_\odot$ model of \cite{cantiello:16}, a magnetic field with a radial component $B_r \gtrsim 10^5 \, {\rm G}$ near the core, or $B_r \gtrsim 3 \times 10^3 \, {\rm G}$ just under the stellar surface of the primary of $\varepsilon$~Lupi would be sufficient to suppress g modes with frequencies of $1 \, {\rm d}^{-1}$. The definitive presence or absence of TEOs would allow us to constrain the internal magnetic field strength as well as providing an important test of the aforementioned theories. 

We do not detect any unambiguous signatures of tidally excited g modes in our photometry. However, Figure \ref{fig:EpsLupTide} shows that these tidally excited g modes are expected to have amplitudes less than $\Delta L/L \lesssim 3 \times 10^{-4}$, just below our detection threshold. Hence, the absence of observed g modes cannot currently be used to constrain the subsurface magnetic fields of the components of $\varepsilon$ Lupi. However, a slighly better photometric precision of $\Delta L/L \sim 5 \times 10^{-5}$ would be sufficient to detect or rule out the presence of tidally excited g modes in $\varepsilon$ Lupi. While data from \textit{TESS} should have the precision necessary \citep{Ricker2015}, unfortunately \el~ falls into a sector gap and will not observed. We will therefore have to wait for  other missions such as PLATO to further our understanding of the role magnetism plays in stellar evolution as \el~ is a special system that can be exploited to this end.

 \section*{Acknowledgments}
 Based on observations obtained at the Canada-France-Hawaii Telescope (CFHT) which is operated by the National Research Council of Canada, the Institut National des Sciences de l'Univers of the Centre National de la Recherche Scientifique of France, and the University of Hawaii. PIONIER was developed by the Universit\'e Grenoble Alpes, the Institut de Plan\'etologie et d'Astrophysique de Grenoble (IPAG), the Agence Nationale pour la Recherche (ANR-06-BLAN-0421, ANR-10-BLAN-0505, ANR-10-LABX56, ANR-11-LABX-13), and the Institut National des Sciences de l'Univers (INSU PNP and PNPS), in collaboration with CEA-LETI based on CNES R\&T funding. This research has made use of the Aspro\footnote{Available at http://www.jmmc.fr/aspro} and SearchCal\footnote{Available at http://www.jmmc.fr/searchcal} services (Jean-Marie Mariotti Center), of the SIMBAD and VIZIER databases (CDS, Strasbourg, France) and of the Astrophysics Data System (NASA). AFJM, MS and GAW acknowledge support from the Natural Sciences and Engineering Research Council of Canada (NSERC). MS acknowledges support from the Annie Jump Cannon Fellowship supported by the University of Delaware and endowed by the Mount Cuba Astronomical Observatory. APo was responsible for image processing and automation of photometric routines for the data registered by BRITE-nanosatellite constellation, and was supported by BKMN grant no. 02/010/BKM18/0136. KZ acknowledges support by the Austrian Fonds zur F{\"o}rderung der wissenschaftlichen Forschung (FWF, project V431-NBL) and the Austrian Space Application Programme (ASAP) of the Austrian Research Promotion Agency (FFG). WW acknowledges support by the Austrian Space Application Programme (ASAP) of the Austrian Research Promotion Agency (FFG). SM acknowledges support by the ERC through SPIRE (grant No.\,647383) and CNES PLATO grant at CEA-Saclay. APi acknowledges support from the NCN grant no. 2016/21/B/ST9/01126. GH acknowledges support by the Polish National Science Center (NCN), grant no. 2015/18/A/ST9/00578.

\bibliography{mn2e}

\end{document}